\documentclass[10pt,journal,twocolumn]{IEEEtran}
\usepackage{amssymb,array,algorithm,algorithmic,amsmath,bm,epsfig,epstopdf,graphicx,hyperref,soul,color,subcaption,stfloats,cite,xcolor,multicol,flushend,lipsum,cuted}
\usepackage[framemethod=tikz]{mdframed}
\graphicspath{{../Figures/}}

\DeclareMathOperator*{\sinc}{sinc}

\DeclareMathOperator*{\otherwise}{otherwise}

\DeclareMathOperator*{\diric}{Diric}
\DeclareMathOperator*{\psinc}{psinc}

\usepackage{tikz}
\newcommand*\circled[1]{\tikz[baseline=(char.base)]{
		\node[shape=circle,draw,inner sep=0.5pt] (char) {#1};}}

\definecolor{cream}{RGB}{45, 45, 45}
\definecolor{newWhite}{RGB}{162, 183, 185}
\definecolor{cream2}{RGB}{235,235,235}

\begin{document}
\markboth{The Hong Kong University of Science and Technology (Guangzhou) \today}{}
\title{Fast Signal Interpolation Through Zero-padding and FFT/IFFT}
\author{
\IEEEauthorblockN{
	Zijun Gong,\,\emph{Member, IEEE}}\\
\thanks{Z. Gong is with the IOT Thrust, HKUST (Guangzhou), Guangzhou, Guangdong 511453, China; and the Department of ECE, HKUST, Hong Kong SAR, China (E-mail: gongzijun@ust.hk).}
}%
\maketitle

\begin{abstract}
Based on the sampling theorem, interpolation should be conducted by employing the sinc functions as the kernels. Inspired by the fact that the discrete Fourier transform (DFT) is sampled from the discrete time Fourier transform, a fast signal interpolation algorithm based on zero-padding and fast Fourier transform (FFT) and inverse FFT (IFFT) is presented. This algorithm gives a good approximate of the ideal interpolation, in spite of the windowing effect. The fundamental difference of this algorithm and the ideal sinc interpolation is unveiled, and shown to be deeply rooted in the connection of the sinc function and the Dirichlet function. 
\end{abstract}


\section{Introduction}

Interpolation is a classic problem in signal procesing. How interpolation should be done is dependent on the characteristics of the signals. There are two classic interpolation methods in signal processing. There are generally two ways for interpolation. The first approach is time-domain interpolation with sinc functions \cite{Schanze1995}, justified by the sampling theorem. Another approach is zero-padding with DFT/IDFT, which makes sense because the DFT is sampled from DTFT. A natural question is: are they the same? In the following, we will briefly review these two methods and identify the connection. 

\subsection{Method I: Interpolation with sinc Function}
Consider a continuous-time\,(CT) signal $x_c(t)$, sampled at an interval of $T_s$ second. The discrete-time sequence is \begin{equation}
 	x[n]=x_c(nT_s),
\end{equation} 
and we can try to recover the original CT signal through interpolation with sinc kernel as
\begin{equation}
	x_r(t)=\sum_{n=-\infty}^{\infty}x[n]\sinc(\pi (t-nT_s)),\label{eqn05}
\end{equation}
where $\sinc(\omega)$ is the sinc function defined as
\begin{equation}
	\sinc(\omega)=\frac{\sin\omega}{\omega}.
\end{equation}
Given that $x_c(t)$ is bandlimited and $T_s$ is small enough, the sampling theorem promises us that the interpolation will be perfect, i.e., $x_r(t)=x_c(t)$\cite{Oppenheim2009DSP}. By doing so we can easily up-sample the signal, i.e., interpolation.

\subsection{Method II: Interpolation with FFT/IFFT}
Consider $x[n]$, a sequence of length $N$, the DTFT is $X(\omega)$ while the $N$-point DFT is $X[k] = X(2\pi k/N)$. If we zero-pad $x[n]$ to a sequence of length $\tilde{N}>N$, i.e., $\tilde{x}[n]$
\begin{equation}	
	\tilde{x}[n]=\left\{
	\begin{array}{ll}
	x[n], & 0\leq n \leq N-1\\
		0, & \otherwise,
	\end{array}
	\right.
\end{equation}
the DFT of $\tilde{x}[n]$ is $\tilde{X}[k] = X(2\pi k/\tilde{N})$. As a result, we can see that $\tilde{X}[k]$ is an up-sampled version of $X[k]$. This is a very classic way for up-sampling, based on zero-padding and FFT/IFFT. 

Method I is for time-domain interpolation, while Method II is for frequency domain. Consider the time-frequency symmetry, one may wonder is it possible to interpolate time-domain signals with zero-padding and FFT/IFFT? Besides, what's the difference between these two methods? Their connection will be unveiled in the following section.

\section{Interpolation of Time-Limited Signals}

Consider the time-frequency symmetry, we should be able to interpolate signals in time-domain in a similar way. However, it's more complicated than that. In this section, we will first revisit the zero-padding based interpolation method in frequency domain, and unveil its fundamental difference with the sinc-based interpolation. Then, we will present a way to revise this algorithm for time-domain interpolation.

\subsection{Zero-padding based Interpolation in Frequency Domain}
To start with, the DFT of a DT sequence $x[n]$ is given by 
\begin{equation}
	\begin{split}
		X(\omega)=& \frac{1}{N}\sum_{n=0}^{N-1}x[n]e^{-jn\omega}\\
		=& \frac{1}{N}\sum_{n=0}^{N-1} \sum_{k=0}^{N-1} X[k]e^{jnk\omega_0}e^{-jn\omega}\\
		=&  \sum_{k=0}^{N-1} X[k]e^{-j\frac{(N-1)}{2}(\omega-k\omega_0)}\underbrace{\frac{\sin N(\omega-k\omega_0)/2}{N\sin(\omega-k\omega_0)/2}}_{\diric(N, \omega-k\omega_0)}
	\end{split}
\end{equation}
where we have the Dirichlet function of order $N$ as
\begin{equation}
	\diric(N,\omega)=\frac{\sin N\omega/2}{N\sin\omega/2}.
\end{equation}
The Dirichlet function is also often referred to as the \emph{periodic sinc} function for two reasons. First, the primary period of the Dirichlet function is very close to the sinc function of the same mainlobe width. Second, the Dirichlet function is periodic with a periodic of $4\pi$ for even $N$ or $2\pi$ for odd $N$.

As we can see, the second way of interpolation is very different from the first one. From the discrete samples, the interpolation actually involves three steps: $(a)$ phase rotation of the discrete sequence; $(b)$ interpolation with Dirichlet kernel; $(c)$ phase adjustment of the new sequence. These three steps can be clearly observed by writing $X(\omega)$ as
\begin{equation}
		 \underbrace{e^{-j\frac{(N-1)}{2}\omega}\underbrace{\sum_{k=0}^{N-1}  \underbrace{X[k]e^{j\frac{(N-1)}{2}k\omega_0}}_{(a)}\diric(N, \omega-k\omega_0)}_{(b)}}_{(c)}.\label{eqn04}
\end{equation}

\begin{figure}[htp]
 	\centering
 	\includegraphics[width=0.5\textwidth]{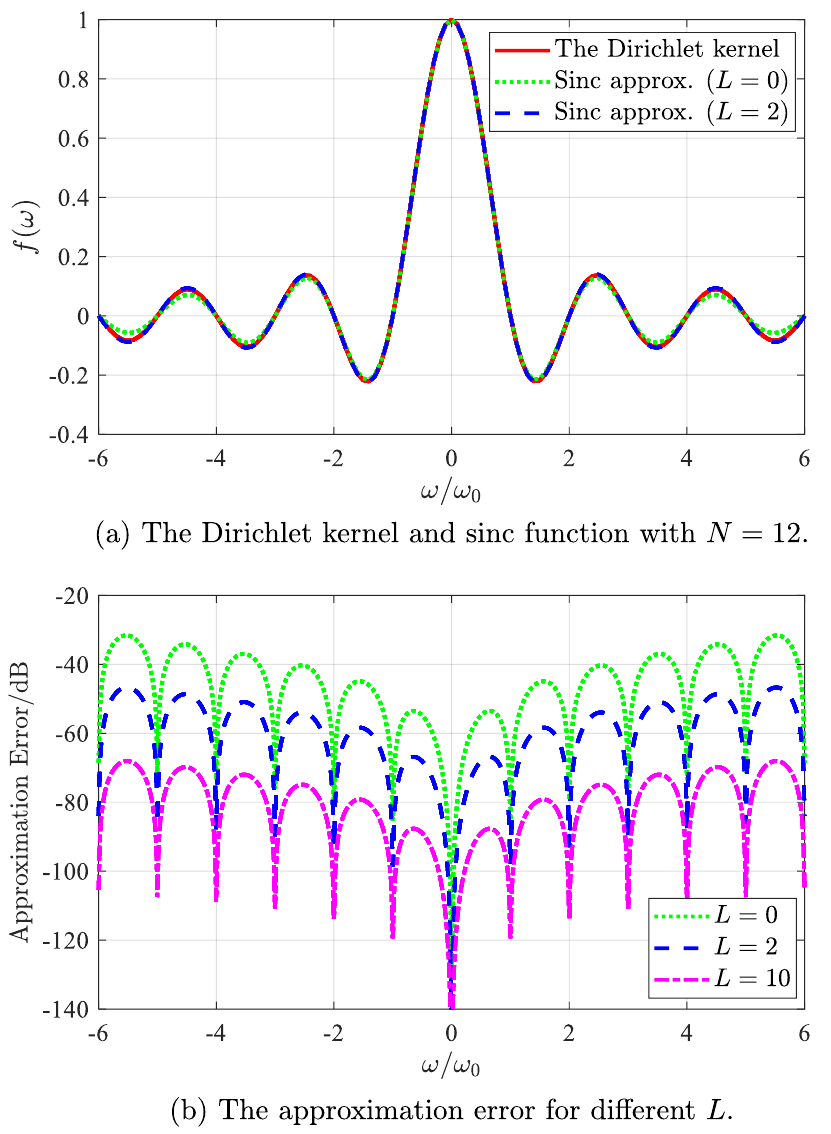}
 	\caption{The sinc function and the Dirichlet kernel with different $L$.}
 	\label{sins}
 \end{figure}

As we can see, the interpolation kernels of the methods are different. The second one corresponds to interpolation with $\frac{N\sin N\omega}{\sin \omega}$, which is always periodic. So the next question is what's the difference between these two ``sinc'' functions? Basically, the Dirichlet function can be obtained through periodic extension of the sinc function with proper phase rotations:
\begin{equation}
 	\diric(N,\omega)=\sum_{l=-\infty}^{\infty}e^{j (N-1)l\pi}\sinc\left(\frac{\omega-2\pi l}{2/N}\right).
\end{equation}
If we define the periodically extended sinc function as 
\begin{equation}
		\psinc(N,L,\omega)=\sum_{l=-L}^{L}e^{j (N-1)l\pi}\sinc\left(\frac{\omega-2\pi l}{2/N}\right),
\end{equation}
we then have 
\begin{equation}
	\diric(N,\omega)=\lim_{L\rightarrow\infty}\psinc(N,L,\omega).
\end{equation}
Apparently, $\psinc$ is the periodized version of the original sinc function, with time-domain overlapping. The discussion in the next sub-section is one way to prove this result. At this point, we will compare them numerically in Fig. \ref{sins}.

%

In Fig. \ref{sins} (a), we can see that $\diric(N,\omega)$ is very close to a scaled sinc function, i.e., $L=0$, and there is an observable fitting discrepancy for large $\omega$. For $L=2$, the discrepancy is already very small. This is consistent with the theoretical conclusion that $\psinc(N,L,\omega)$ is getting closer to $\diric(N,\omega)$ when $L$ grows. In (b), the difference between $\diric(N,\omega)$ and periodically extended sinc function is compared for different $L$, i.e., $20\lg(|\diric(N,\omega)-\psinc(N,L,\omega)|)$. As we can see, the difference decreases with $L$, and for $L=10$, the difference is already smaller than -60\,dB. We can also see that the discrepancy tends to be smaller for $\omega$ close to 0. This can be explained by the fact that $\sin(\omega)\approx \omega$ for $|\omega|\ll 1$.
 
\subsection{Interpolation of Time-Limited Signals}
From the previous discussions, we can see that the two classic interpolation methods are fundamentally different. Nonetheless, the zero-padding based interpolation method involves interpolation with the Dirichlet function, which is closely related to the sinc function. Therefore, it might be possible to revise the second method in a way and use it for time-domain signal interpolation, which is the focus of this sub-section.

Consider $x_c(t)$ compactly supported for $t\in[0,T]$, and sample it with a period of $T_s$. Without loss of generality, let $T=(N-1)T_s$ and we have
 \begin{equation}
 	x[n]=x_c(nT_s).
 \end{equation}
Suppose the Fourier transform of $x_c(t)$ is $X_c(\Omega)$, the DTFT of $x[n]$ is given as
 \begin{equation}
 	\begin{split}
 	X(\omega)=&\frac{1}{T_s}\sum_{l=-\infty}^{\infty} X_c\left(\frac{\omega - 2\pi l}{T_s}\right).\\
 	\end{split}
 	\label{eqn03}
 \end{equation}
 Due to the limited time span of the signal $x_c(t)$, we can use a discrete sequence/sampled version to perfectly recover $X_c(\Omega)$. To start with, we shift $x_c(t)$ in time so that it's centered at $t=0$, i.e., $x_c(t-T/2)$. The corresponding Fourier transform is $X_c(\Omega)e^{j\Omega (N-1)T_s/2}$. Then suppose we sample the spectrum with a period of $\Omega_s=2\pi/(NT_s)\leq 2\pi/T$. Based on the sampling theorem, we then have \eqref{eqn01}, which states that the Fourier transform of the signal can be written as a weighted sum of shifted sinc functions.
\newcounter{TempEqCnt} 
\setcounter{TempEqCnt}{\value{equation}}
\setcounter{equation}{12} 
\begin{figure*}[htb]
	\begin{equation}
		\begin{split}
			X_c(\Omega)=&e^{-j\Omega(N-1)T_s/2}\underbrace{\sum_{r=-\infty}^{\infty} X_c(r\Omega_s)e^{jk\Omega_s(N-1)T_s/2}\frac{\sin\pi(\Omega-r\Omega_s)/\Omega_s}{\pi(\Omega-r\Omega_s)/\Omega_s}}_{X_c(\Omega)e^{j\Omega(N-1)T_s/2}}\\
			=&\sum_{r=-\infty}^{\infty} X_c(r\Omega_s)e^{j(\Omega-r\Omega_s)(N-1)T_s/2}\sinc(\pi(\Omega/\Omega_s-r))\\
		\end{split}\label{eqn01}
	\end{equation}
	\hrule
\end{figure*}

Let $\omega_0=\Omega_sT_s=2\pi/N$, and we have \eqref{eqn02}. Step (1) is a direct result of \eqref{eqn03} and \eqref{eqn01}. In step (3), $r$ is replaced by $k+rN$, and $k+rN$ ranges from $-\infty$ to $\infty$. 
\begin{figure*}
 \begin{equation}
 	\begin{split}
 	X(\omega)\overset{(1)}{=}&\frac{1}{T_s}\sum_{l=-\infty}^{+\infty} \sum_{r=-\infty}^{+\infty} X_c(r\Omega_s)e^{-j (N-1)(\omega-r\omega_0-lN\omega_0)/2}\sinc\left(\frac{\omega}{2}N-r\pi-lN\pi\right)\\
 		\overset{(2)}{=} &\frac{1}{T_s} \sum_{l=-\infty}^{+\infty} \sum_{r=-\infty}^{+\infty} X_c(r\Omega_s)e^{-j (N-1)(\omega-r\omega_0)/2}\sinc\left(\frac{\omega-2\pi l}{2/N}-r\pi\right)\\
 		\overset{(3)}{=} &\frac{1}{T_s}\sum_{l=-\infty}^{+\infty} \sum_{r=-\infty}^{+\infty} \sum_{k=0}^{N-1} X_c((k+rN)\Omega_s)e^{-j (N-1)(\omega-k\omega_0-(l+r)N\omega_0 )/2}\sinc\left(\frac{\omega-2\pi (l+r)}{2/N}-k\pi\right)\\
 		\overset{(4)}{=} &\sum_{k=0}^{N-1}\underbrace{\frac{1}{T_s}\sum_{r=-\infty}^{+\infty}X_c((k+rN)\Omega_s)}_{\tilde{X}_c[k]}e^{-j (N-1)(\omega-k\omega_0 )/2}\sum_{l=-\infty}^{+\infty} e^{j (N-1)l\pi}\sinc\left(\frac{\omega-k\omega_0-2\pi l}{2/N}\right)\\
 	\end{split}\label{eqn02}
 \end{equation}
 \hrule
\end{figure*}

 Meanwhile, suppose $X[k]$ is the DFT of $x[n]$, and we have
 \begin{equation}
 	\begin{split}
 		X(\omega)=& \frac{1}{N}\sum_{n=0}^{N-1} x[n]e^{-jn\omega}\\
 		=& \frac{1}{N}\sum_{n=0}^{N-1} \sum_{k=0}^{N-1} X[k]e^{jnk\omega_0}e^{-jn\omega}\\
 		=&  \frac{1}{N}\sum_{k=0}^{N-1} X[k]\sum_{n=0}^{N-1}e^{-jn(\omega-k\omega_0)}\\
 		=&  \sum_{k=0}^{N-1} X[k]e^{-j\frac{(N-1)(\omega-k\omega_0)}{2}}\diric(N,\omega-k\omega_0).
 	\end{split}\label{eqn05}
 \end{equation}
 The DFT is related to the DTFT through $X[k']=X(k'\omega_0)$, and from \eqref{eqn03} we have
 \begin{equation}
 X[k]= \tilde{X}_c[k].
 \end{equation}
By comparing \eqref{eqn02} and \eqref{eqn05}, note that they must be equal for any $x_c(t)$, we can conclude that 
 \begin{equation}
 	\frac{\sin N\omega/2}{N\sin \omega/2}=\sum_{l=-\infty}^{\infty}e^{j (N-1)l\pi}\sinc\left(\frac{\omega-2\pi l}{2/N}\right).
\label{eqn06}
 \end{equation}

From here, the interpolation algorithm is clear. For a sequence $x[n]$ of length $N$, suppose we want to increase the number of samples by a factor of 2. The new sequence can be obtained in three steps as follows.
\begin{itemize}
	\item [{\circled{1}}] We multiply the sequence with a phase rotation: 
	\begin{displaymath}
		x_1[n] = x[n]e^{-j(N-1)n\pi/N}.
	\end{displaymath}
	\item [{\circled{2}}] After $N$-point IFFT on $x_1[n]$, followed by $2N$-point FFT, the new sequence is $x_2[n]$ of length $2N$.
	\item [{\circled{3}}] Add phase adjustment to the new sequence:
	\begin{displaymath}
		x_3[n]=x_2[n]e^{j(N-1)(2n\pi/2N)/2}=x_2[n]e^{j(N-1)n\pi/(2N)}.
	\end{displaymath}
\end{itemize}
The phase adjustments in step \circled{1} and \circled{3} can cancel the phase rotations in step $(a)$ and $(c)$ in \eqref{eqn04}, respectively. The obtained sequence is thus related to the original sequence as
\begin{equation}
	\begin{split}
	x_3[n]=&\sum_{k=0}^{N-1}  x[n]\diric(N, n\pi/N-k\omega_0)\\
	      =&\sum_{k=0}^{N-1}  x[n]\diric(N, n\pi/N-2\pi k/N)\\
	      =&\sum_{k=0}^{N-1}  x[n]\diric(N, (n-2k)\pi/N)\\
	      \approx&\sum_{k=0}^{N-1}  x[n]\sinc\left(\frac{(n-2k)\pi/N}{2/N}\right)\\
	      =&\sum_{k=0}^{N-1}  x[n]\sinc\left(\pi(n/2-k)\right).
	\end{split}
\end{equation}
The approximation is based on the resemblance between the sinc function and the Dirichlet function, for $L=0$ in Fig. \ref{sins}. This justifies the effectiveness of the proposed method based on zero-padding and FFT/IFFT. 


\section{Conclusions}
The sinc function is common see as a result of Fourier transform of a CT rectangular function. If we sample the rectangular function and do DFT, the Dirichlet function appears. Because the rectangular function is not bandlimited, aliasing will happen when we sample it. This is another way to understand why the Dirichlet function is a periodic extension of the sinc function in \eqref{eqn06}.

Based on this observation, we can use the zero-padding and FFT/IFFT based algorithm for time-domain signal interpolation. What is happening underneath is that we are replacing the sinc function with the Dirichlet function as the interpolation kernel. Such an approach provides accurate reconstruction of the time domain signal.




\bibliographystyle{ieeetr}
\bibliography{signalInterpolation}
\end{document}